\documentclass[runningheads]{llncs}
\usepackage[T1]{fontenc}

\usepackage{graphicx}
\graphicspath{{figures/}{exppic/}} 
\usepackage{float}        
\usepackage{wrapfig}

\usepackage{subcaption}

\usepackage[font=small,labelfont=bf]{caption}

\usepackage{multirow}
\usepackage{amsmath,amssymb}
\usepackage{algorithm}
\usepackage{algorithmic}

\usepackage{xurl}      
\usepackage{url}
\usepackage[
    backend=biber,
    style=ieee,
    citestyle=numeric-comp,
    sorting=none,
    maxnames=3,       
    minnames=3,       
    giveninits=true   
]{biblatex}
\AtBeginBibliography{\footnotesize}

\usepackage[colorlinks=true,
            linkcolor=blue,
            citecolor=blue,
            urlcolor=blue,
            anchorcolor=blue]{hyperref}

\usepackage[nameinlink,capitalise,noabbrev]{cleveref}

\usepackage{marvosym}   
\usepackage{microtype} 
\begin{document}
\title{TD3-Sched: Learning to Orchestrate Container-based Cloud–Edge Resources via Distributed Reinforcement Learning}
\titlerunning{TD3-Sched: DRL for Container Orchestration in Cloud-Edge}
%

\author{
  Shengye Song\inst{1,2} \and
  Minxian Xu\inst{2}{\Letter} 
  \and
  Kan Hu\inst{2} \and
  Wenxia Guo\inst{3}{\Letter} \and
  Kejiang Ye\inst{2}
}
\authorrunning{S. Song et al.}

\institute{
  Southern University of Science and Technology, Shenzhen, China, 518055\\
  \and
  Shenzhen Institutes of Advanced Technology, Chinese Academy of Sciences, Shenzhen, China, 518055\\
  \and
  Nation Key Laboratory of Electromagnetic Energy, Naval University of Engineering, Wuhan, Hubei Province, China, 430030\\
}

\maketitle              
\vspace{-0.6cm}
\begin{abstract}

Resource scheduling in cloud-edge systems is challenging as edge nodes run latency-sensitive workloads under tight resource constraints, while existing centralized schedulers can suffer from performance bottlenecks and user experience degradation. To address the issues of distributed decisions in cloud-edge environments, we present \textbf{TD3-Sched}, a distributed reinforcement learning (DRL) scheduler based on \emph{Twin Delayed Deep Deterministic Policy Gradient (TD3)} for continuous control of CPU and memory allocation, which can achieve optimized decisions for resource provisioning under dynamic workloads. On a realistic cloud-edge testbed with SockShop application and Alibaba traces, TD3-Sched achieves reductions of 17.9\% to 38.6\% in latency under same loads compared with other reinforcement-learning and rule-based baselines, and 16\% to 31.6\% under high loads. TD3-Sched also shows superior \emph{Service Level Objective (SLO)} compliance with only 0.47\% violations. These results indicate faster convergence, lower latency, and more stable performance while preserving service quality in container-based cloud-edge environment compared with the baselines.
\end{abstract}
\vspace{-1cm}
\vspace{-0.3cm}
\section{Introduction}
\label{section:1}

\vspace{-0.2cm}

Cloud-edge computing has emerged as a critical paradigm for delivering low-latency services in distributed environments. However, the inherent heterogeneity of cloud-edge architectures poses significant challenges for resource orchestration with the consideration of container life-cycles, where edge nodes operate with constrained resources while cloud centers provide abundant computational power but suffer from network latency \cite{erlang2024,topfull2024}.
Cloud-edge resource orchestration faces challenges including resource constraints at edge nodes, real-time decision-making for latency-sensitive applications, and complexity in workload migration across heterogeneous environments.
To make our objective explicit, we formulate the cloud–edge resource scheduling problem as: given a set of containerized services running on heterogeneous cloud and edge nodes, determine CPU and memory allocations in real time so as to minimize latency, satisfy SLOs, and maximize resource efficiency under dynamic workloads and migration constraints.

Current resource scheduling approaches face significant limitations: static allocation strategies fail to adapt to changing patterns, while rule-based systems lack sophistication for multi-dimensional optimization \cite{deepscaling2024}. Existing Kubernetes schedulers focus on container placement but lack mechanisms for dynamic resource adjustment and latency optimization \cite{wang2024autothrottle}. Machine learning approaches often focus on single objectives, neglecting the critical importance of latency optimization in edge computing scenarios \cite{graf2024}. RL has emerged as a promising approach for addressing complex decision-making challenges in resource management, but most existing applications focus on cloud-only environments or single-objective optimization problems, leaving the application to multi-objective optimization in cloud-edge environments largely unexplored.

To address these issues, we present TD3-Sched, a DRL resource scheduling system for cloud-edge computing environments. TD3-Sched formulates resource scheduling as a continuous control problem using the TD3 algorithm, which addresses overestimation bias and training instability issues through dual Q-networks and target policy smoothing. The system operates on a 4N-dimensional state space representing CPU utilization, memory utilization, response latency, and Queries Per Second (QPS) indicators, and produces 2N-dimensional actions for CPU and memory allocation decisions. The reward function balances four competing objectives: latency penalty, resource waste penalty, SLO satisfaction reward, and migration cost penalty, with carefully tuned weights prioritizing latency optimization. The architecture integrates comprehensive state monitoring, TD3-based RL with dual Q-networks and target policy smoothing, and Kubernetes API integration for resource allocation execution, similar to recent approaches in production cloud systems \cite{aware2023,cilantro2023}.

Under the Kubernetes\footnote{https://kubernetes.io/}-based prototype system with realistic Sock Shop application and Alibaba workloads traces, our evaluation against Deep Q-Network (DQN) \cite{zhai2025joint}, Deep Deterministic Policy Gradient (DDPG) \cite{xie2024deep}, and BaseK \cite{poulton2023kubernetes} demonstrates TD3-Sched's superior performance, achieving 75.2ms average latency under normal loads and maintaining excellent SLO compliance with only 0.47\% violation rate under high loads conditions.
Our \textbf{key contributions} are summarized as follows: 
 \begin{itemize}
   \item We propose an optimized multi-objective DRL model tailored for diverse resource conditions, ensuring SLO satisfaction while enhancing latency, resource utilization, and migration cost efficiency;
 \item We propose a DRL algorithm specifically designed for container-based cloud–edge resource orchestration, addressing the inherent challenges of heterogeneous and distributed environments by employing dual Q-networks;
   \item We present an integrated system architecture that seamlessly extends Kubernetes with DRL capabilities, featuring real-time state monitoring across multiple nodes, intelligent resource allocation decisions, and results are validated with realistic application and traces.
\end{itemize} 

\section{Background and Related Work}

\label{section:2}

\vspace{-0.2cm}
\subsection{Challenges in Container-based Cloud-Edge Computing }
Cloud–edge computing has emerged as a promising paradigm to meet the increasing demands of low-latency, data-intensive, and geographically distributed applications \cite{geoscale2024}. By integrating the abundant computational capacity of centralized cloud data centers with the proximity advantages of edge nodes, this architecture enables improved service responsiveness and optimized resource utilization. Cloud centers excel at handling compute-intensive workloads at scale, while edge nodes reduce network latency by processing requests closer to end users, thereby enhancing the user experience for time-critical services.

Container based cloud–edge environments face challenges in resource scheduling due to resource constraints at edge nodes and the need for latency-sensitive applications to maintain service quality \cite{deepscaling2024,jiagu2024}. Furthermore, container-based deployments add another layer of complexity, as scheduling must consider container startup overheads, image distribution delays, and runtime performance variability across heterogeneous hardware and dynamic workloads.

The dynamic and heterogeneous nature of cloud–edge environments further complicates scheduling decisions. Workloads can fluctuate unpredictably due to time-varying user demands, mobility patterns, and network congestion, requiring schedulers to make near-real-time decisions that adapt to changing conditions \cite{oceanus2025,song2025c}. In addition, heterogeneous hardware configurations, diverse network topologies, and varying application QoS requirements mean that resource management must simultaneously account for computation, communication, and migration costs. This interplay of constraints and objectives transforms container scheduling into a high-dimensional optimization problem, where balancing efficiency, latency, and reliability remains a significant research challenge \cite{sigmaos2024}.
\vspace{-0.3cm}
\subsection{Limitations of the Existing Work}
The current related work in container orchestration in cloud-edge environment can be mainly divided into two categories: 

\textbf{Reinforcement Learning in Resource Management.} Reinforcement learning has emerged as a promising approach for addressing complex decision-making challenges in resource management systems. DQN has been applied to discrete resource allocation problems, utilizing experience replay and target networks to improve learning stability \cite{aware2023,jiagu2024}. However, DQN's discrete action space limits its applicability to continuous resource allocation scenarios. DDPG algorithms offer continuous control capabilities but suffer from overestimation bias and training instability. The TD3 algorithm addresses these limitations through dual Q-networks, target policy smoothing, and delayed policy updates, providing more stable and reliable learning for continuous control problems \cite{cilantro2023}.

\textbf{Multi-Objective Optimization and Cloud-Edge Scheduling.}
Resource scheduling in cloud-edge environments involves multiple competing objectives that must be balanced simultaneously \cite{geoscale2024,oceanus2025}. Traditional approaches use weighted sum methods or Pareto optimization techniques, but these methods are sensitive to weight selection and can't capture complex trade-offs \cite{deepscaling2024,starburst2024}. Recent work has explored multi-objective reinforcement learning for resource management, but these approaches often focus on specific domains or simplified scenarios \cite{graf2024,erlang2024}.

The integration of reinforcement learning with container orchestration platforms has shown promise, but existing implementations often lack sophisticated mechanisms for dynamic resource adjustment and latency optimization in cloud-edge scenarios \cite{wang2024autothrottle,topfull2024}. TD3-Sched addresses these limitations by extending TD3 algorithms to cloud-edge resource scheduling, introducing a comprehensive reward function that balances multiple competing objectives, and providing an integrated system architecture that extends Kubernetes with reinforcement learning capabilities while maintaining production deployment reliability \cite{sigmaos2024,oppertune2024}.

\textbf{Limitations of Current Approaches.}
Current resource scheduling approaches in container-based cloud-edge environments suffer from several fundamental limitations. Static resource allocation strategies, commonly used in traditional cloud deployments, allocate fixed amounts of CPU and memory resources to containers based on initial estimates \cite{starburst2024}. While simple to implement, these approaches fail to adapt to changing workload patterns and often result in either resource waste or performance degradation.
Rule-based scheduling systems attempt to address this limitation by implementing predefined policies for resource adjustment, but these policies are typically based on simple thresholds and lack the sophistication needed to handle complex, multi-dimensional optimization problems \cite{deepscaling2024,oppertune2024}. Existing Kubernetes-based scheduling solutions, while effective for cloud environments, face significant limitations when applied to cloud-edge scenarios \cite{wang2024autothrottle,cilantro2023}. The default Kubernetes scheduler primarily focuses on container placement decisions based on resource requests and node capacity, but it lacks the sophisticated mechanisms needed to handle dynamic resource adjustment and latency optimization.

Machine learning-based approaches have shown promise in addressing some of these limitations, but existing implementations often focus on single objectives such as resource utilization or energy efficiency, neglecting the critical importance of latency optimization in edge computing scenarios \cite{graf2024,aware2023}. Reinforcement learning has emerged as a promising approach for addressing the complex decision-making challenges in resource management systems, enabling systems to learn optimal policies through interaction with the environment. However, most existing applications of reinforcement learning in resource management focus on cloud-only environments or single-objective optimization problems, leaving the application to multi-objective optimization in cloud-edge environments largely unexplored. The key advantage of distributed reinforcement learning lies in its ability to scale across multiple agents or nodes, enabling parallelized decision-making and reducing latency. By separating the training process  from execution on distributed across edge and cloud nodes, DRL allows for real time execution of resource allocation decisions, which is crucial in cloud-edge environments where workloads are dynamic and latency sensitive. This distributed approach also enables localized learning and decision-making, reducing the burden on a central agent and improving overall system responsiveness.

\vspace{-0.5cm}
\section{TD3-Sched: System Architecture Design}
\label{section:3}
\begin{figure}[t]
\centering
\includegraphics[width=0.95\textwidth]{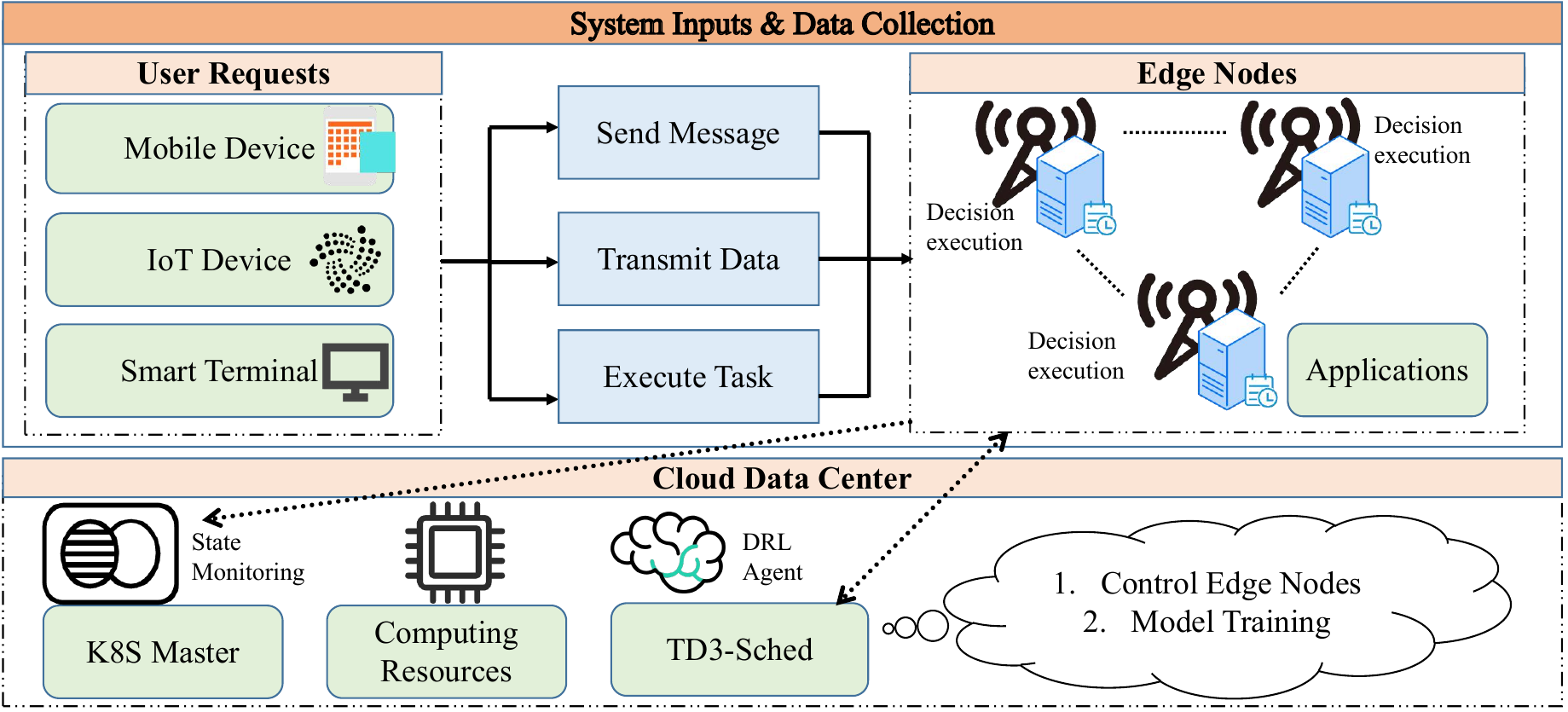}
\vspace{-0.3cm}
\caption{TD3-Sched System Architecture. The system integrates edge nodes with cloud data centers through a reinforcement learning-based scheduling framework, enabling intelligent resource management across heterogeneous distributed environments.}
\vspace{-0.6cm}
\label{fig:system_architecture}
\end{figure}

\vspace{-0.2cm}
Figure \ref{fig:system_architecture} illustrates the TD3-Sched system architecture, which leverages DRL to enable resource orchestration across distributed edge nodes and cloud data centers. Unlike traditional static and centralized approaches, our system employs a distributed architecture by distributing decision-making responsibilities to different nodes that enables intelligent resource orchestration across multiple edge nodes and cloud centers through DRL coordination.

The framework formulates resource scheduling as a continuous control problem, where DRL agents learn optimal resource allocation policies through direct interaction with the distributed environment. This enables handling multi-dimensional optimization challenges in cloud-edge environments, balancing latency minimization, resource utilization, and system stability across edge nodes and cloud data centers through distributed resource management.

The architecture consists of three core components. The distributed state monitoring module collects system metrics including CPU utilization, memory utilization, response latency, and QPS metrics from multiple cloud data centers and edge nodes, establishing a comprehensive view of the distributed system state. The DRL agents implement the TD3 algorithm, operating on a 4N-dimensional state space and producing 2N-dimensional actions representing CPU and memory allocation decisions for services across the distributed infrastructure. The distributed decision execution module translates these actions into concrete resource allocations through the Kubernetes API, enabling coordinated resource management across multiple nodes.

The reward function balances four objectives: latency penalty, resource waste penalty, SLO satisfaction reward, and migration cost penalty, with weights carefully tuned for distributed cloud-edge environments. The integration extends the Kubernetes scheduling framework with DRL capabilities, maintaining compatibility while adding advanced resource management features for distributed infrastructure. The system follows a two-phase deployment approach: offline training using historical data from distributed nodes, followed by online deployment with continuous learning. Real-time adaptation enables quick response to changing workload patterns and resource availability across the distributed cloud-edge infrastructure, while maintaining low latency for resource allocation decisions. This DRL architecture addresses the fundamental challenges of cloud-edge resource scheduling through intelligent, adaptive decision-making while ensuring production deployment reliability and scalability across heterogeneous distributed environments.

\vspace{-0.5cm}

\section{TD3-Sched: Algorithm Design and Implementation}
\label{section:4}
\begin{figure}
\centering
\includegraphics[width=0.95\textwidth]{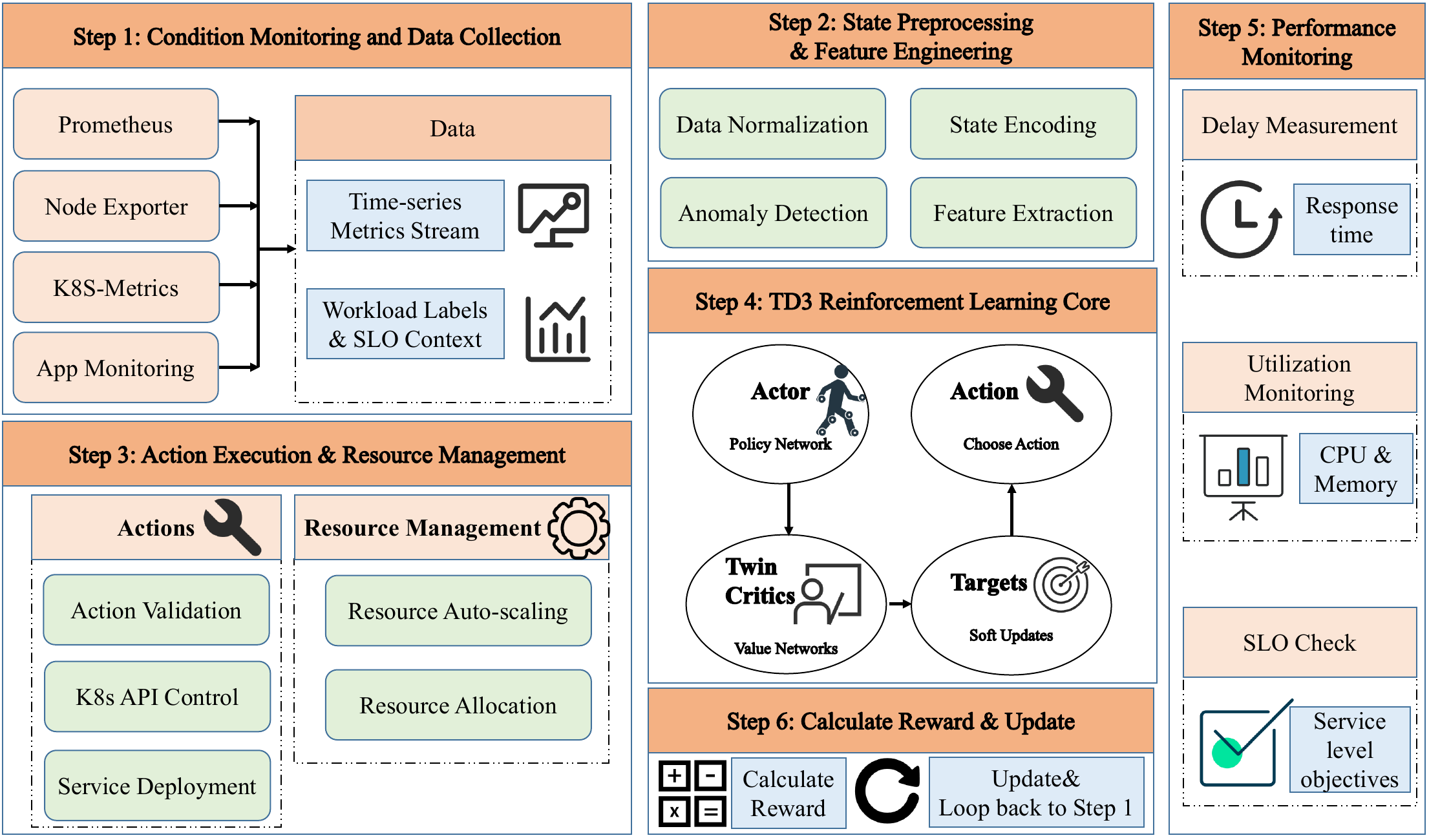}
\caption{TD3-Sched System Workflow. The system implements a continuous learning process that monitors system conditions, processes state information, makes intelligent decisions through reinforcement learning, executes resource allocations, and learns from feedback to improve future decisions.}
\label{fig:td3sched_workflow}
\vspace{-0.1cm}
\end{figure}

\vspace{-0.2cm}
\textbf{Algorithm Overview.}
Figure \ref{fig:td3sched_workflow} illustrates the complete TD3-Sched distributed system workflow, which implements a continuous closed-loop process for intelligent resource scheduling across distributed cloud-edge environments.

The distributed workflow begins with comprehensive data collection from multiple sources across distributed nodes including Prometheus\footnote{https://prometheus.io/}, Node Exporter, Kubernetes metrics, and application monitoring systems. This raw data is then preprocessed through normalization, anomaly detection, feature extraction, and state encoding to create a 4N-dimensional state vector representing the current distributed system condition. The TD3-Sched reinforcement learning core processes this state information to generate optimal resource allocation decisions, which are then executed through the Kubernetes API across multiple nodes to adjust CPU and memory allocations for each service.

We formulate the resource scheduling problem in cloud-edge environments as a Markov Decision Process (MDP) defined by the tuple $(\mathcal{S}, \mathcal{A}, \mathcal{P}, \mathcal{R}, \gamma)$, where $\mathcal{S}$ is the state space, $\mathcal{A}$ is the action space, $\mathcal{P}$ is the transition probability function, $\mathcal{R}$ is the reward function, and $\gamma \in [0, 1]$ is the discount factor.

The system state $s_t \in \mathcal{S}$ at time step $t$ is defined as:
\begin{align}
s_t = [\mathbf{c}_t, \mathbf{m}_t, \mathbf{l}_t, \mathbf{q}_t].
\label{equ:state_space}
\end{align}
where $\mathbf{c}_t$, $\mathbf{m}_t$, $\mathbf{l}_t$, and $\mathbf{q}_t$ represent CPU utilization, memory utilization, response latency, and QPS for each service respectively. Each state component is normalized: $s_t \in [0, 1]^{4N}$.

The agent's action $a_t \in \mathcal{A}$ represents the resource allocation strategy:
\begin{align}
a_t = [\mathbf{cpu}_t, \mathbf{mem}_t].
\label{equ:action_space}
\end{align}
where $\mathbf{cpu}_t$ and $\mathbf{mem}_t$ represent CPU and memory allocation decisions for each service, constrained by:
\begin{align}
cpu_i^t \in [0.1, 2.0], \quad mem_i^t \in [64, 2048].
\label{equ:action_constraints}
\end{align}

\textbf{Algorithm Implementation.}
TD3-Sched is an optimized DRL algorithm based on TD3, designed for resource scheduling in distributed cloud–edge environments. Its key innovation lies in enabling stable continuous resource allocation through domain-specific modifications, while simultaneously optimizing latency, ensuring SLO compliance, and improving resource utilization under heterogeneous and dynamic network conditions.

The reward function balances multiple competing objectives critical for cloud-edge environments:
\begin{align}
R(s_t, a_t) = \alpha \cdot R_l(s_t) + \beta \cdot R_r(a_t) + \lambda \cdot R_s(s_t) + \mu \cdot R_m(a_t, a_{t-1}).
\label{equ:reward_function}
\end{align}
The components address different aspects: 

1) Latency penalty $R_l(s_t) = -\sum_{i=1}^{N} \max(0, l_i^t - l_{target})$ penalizes response times exceeding SLO targets.

2) Resource waste $R_r(a_t) = -\sum_{i=1}^{N} \left(\frac{cpu_i^t - c_i^t}{cpu_i^t} + \frac{mem_i^t - m_i^t}{mem_i^t}\right)$ discourages over-allocation.

3) SLO satisfaction $R_s(s_t) = \sum_{i=1}^{N} \mathbb{I}(l_i^t \leq l_{target})$  meeting SLOs. 

4)Migration cost $R_m(a_t, a_{t-1}) = -\sum_{i=1}^{N} \left(|cpu_i^t - cpu_i^{t-1}| + |mem_i^t - mem_i^{t-1}|\right)$ prevents frequent large changes. 

The weights are carefully tuned: $\alpha = 0.5, \beta = 0.1, \lambda = 0.2, \mu = 0.1$, prioritizing latency optimization for edge computing scenarios.

TD3-Sched implements the core TD3 algorithm with domain-specific modifications for cloud-edge resource scheduling. The algorithm maintains two Q-networks $Q_{\phi_1}$ and $Q_{\phi_2}$ to reduce overestimation bias, and a single policy network $\mu_\theta$ for action selection. Here, $\phi_1$ and $\phi_2$ represent the parameters of the two critic networks, and $\theta$ represents the parameters of the actor network.

The target policy smoothing mechanism adds clipped noise to target actions to reduce variance in policy updates:
\begin{align}
a' = \mu_{\theta'}(s') + \text{clip}(\mathcal{N}(0, \sigma^2), -c, c),
\label{equ:target_smoothing}
\end{align}
where $a'$ is the target action, $\mu_{\theta'}(s')$ is the target policy output, $\mathcal{N}(0, \sigma^2)$ is Gaussian noise with standard deviation $\sigma = 0.2$, and $\text{clip}(\cdot, -c, c)$ clips the noise to the range $[-c, c]$ with $c = 0.5$. This mechanism prevents the policy from exploiting errors in the value function.

The temporal difference target for Q-value learning is computed using the minimum of two Q-values to reduce overestimation bias:
\begin{align}
y = r + \gamma \cdot \min_{i=1,2} Q_{\phi_i'}(s', a') \cdot (1 - \text{done}),
\label{equ:td_target}
\end{align}
where $y$ is the TD target, $r$ is the immediate reward, $\gamma = 0.99$ is the discount factor, $Q_{\phi_i'}(s', a')$ is the Q-value from the $i$-th target critic network, and $\text{done}$ is a binary indicator for episode termination. The minimum operation helps reduce overestimation bias that commonly occurs in Q-learning.

The critic networks are updated by minimizing the squared TD error:
\begin{align}
\mathcal{L}(\phi_i) = \mathbb{E}_{(s,a,r,s') \sim \mathcal{D}} \left[ \left( Q_{\phi_i}(s, a) - y \right)^2 \right],
\label{equ:critic_loss}
\end{align}
where $\mathcal{L}(\phi_i)$ is the loss function for the $i$-th critic network, $\mathbb{E}_{(s,a,r,s') \sim \mathcal{D}}$ denotes expectation over transitions sampled from the experience replay buffer $\mathcal{D}$, and $Q_{\phi_i}(s, a)$ is the Q-value estimate from the $i$-th critic network.

The policy network is updated less frequently (every 2 steps) to improve stability, following the delayed policy update mechanism:
\begin{align}
\nabla_\theta J(\theta) = \mathbb{E}_{s \sim \mathcal{D}} \left[ \nabla_a Q_{\phi_1}(s, a) \big|_{a=\mu_\theta(s)} \nabla_\theta \mu_\theta(s) \right],
\label{equ:policy_gradient}
\end{align}
where $\nabla_\theta J(\theta)$ is the policy gradient, $\nabla_a Q_{\phi_1}(s, a)$ is the gradient of the first critic network with respect to actions, and $\nabla_\theta \mu_\theta(s)$ is the gradient of the policy network with respect to its parameters.

Target networks are updated using soft updates with coefficient $\tau = 0.005$:
\begin{align}
\theta' \leftarrow \tau \theta + (1-\tau) \theta', \quad \phi_i' \leftarrow \tau \phi_i + (1-\tau) \phi_i',
\label{equ:soft_update}
\end{align}
where $\theta'$ and $\phi_i'$ are the target network parameters, and the soft update ensures stable learning by gradually incorporating changes from the main networks.

The exploration strategy employs a decaying noise schedule to balance exploration and exploitation:
\vspace{-0.2cm}
\begin{align}
\sigma_{explore}(t) = \sigma_{init} \cdot \exp(-t/\tau_{decay}),
\label{equ:exploration_noise}
\end{align}
where $\sigma_{explore}(t)$ is the exploration noise at time step $t$, $\sigma_{init} = 0.3$ is the initial noise standard deviation, and $\tau_{decay} = 1000$ is the decay time constant. This schedule gradually reduces exploration as the agent learns optimal policies.

TD3-Sched employs an Actor-Critic architecture with dual Q-networks to reduce overestimation bias, a critical improvement over traditional single-critic approaches. The Actor network $\mu_\theta(s)$ maps states to continuous actions using a three-layer architecture with Rectified Linear Unit (ReLU) activations in hidden layers and hyperbolic tangent (tanh) in the output layer, ensuring bounded action outputs. The Critic networks $Q_{\phi_i}(s,a)$ estimate state-action values using the same architecture but with linear output activation, providing stable value estimation for the continuous action space.

\begin{algorithm}[t]
\caption{TD3-Sched Training Algorithm}
\label{alg:td3sched}
\begin{algorithmic}[1]
\REQUIRE Environment env, TD3-Sched agent agent, replay buffer buffer
\ENSURE Trained TD3-Sched agent
\STATE Initialize Actor network $\mu_\theta$ and Critic networks $Q_{\phi_1}, Q_{\phi_2}$
\STATE Initialize target networks $\mu_{\theta'}, Q_{\phi_1'}, Q_{\phi_2'}$
\STATE Initialize replay buffer $\mathcal{D}$
\FOR{episode = 1 to M}
    \STATE $s_t \leftarrow$ env.reset()
    \FOR{step = 1 to T}
        \STATE $a_t \leftarrow \mu_\theta(s_t) + \mathcal{N}(0, \sigma_{explore})$
        \STATE $s_{t+1}, r_t, done \leftarrow$ env.step($a_t$)
        \STATE Store transition $(s_t, a_t, r_t, s_{t+1}, done)$ in $\mathcal{D}$
        
        \IF{$|\mathcal{D}| > batch\_size$}
            \STATE Sample batch $(s, a, r, s', done)$ from $\mathcal{D}$
            \STATE $a' \leftarrow \mu_{\theta'}(s') + \text{clip}(\mathcal{N}(0, \sigma^2), -c, c)$
            \STATE $y \leftarrow r + \gamma \cdot \min_{i=1,2} Q_{\phi_i'}(s', a') \cdot (1 - done)$
            \STATE Update $Q_{\phi_i}$ by minimizing $(Q_{\phi_i}(s, a) - y)^2$
            
            \IF{step \% policy\_freq == 0}
                \STATE Update $\mu_\theta$ by maximizing $Q_{\phi_1}(s, \mu_\theta(s))$
                \STATE Soft update target networks
            \ENDIF
        \ENDIF
        
        \STATE $s_t \leftarrow s_{t+1}$
        \IF{done} \STATE BREAK \ENDIF
    \ENDFOR
\ENDFOR
\end{algorithmic}

\end{algorithm}

The network architecture is specifically designed for cloud-edge resource scheduling, with input layers sized to accommodate the 4N-dimensional state space and output layers configured for the 2N-dimensional action space. This design enables efficient processing of heterogeneous resource metrics while maintaining computational efficiency for real-time decision making.

Algorithm \ref{alg:td3sched} presents the core TD3-Sched training procedure. The algorithm incorporates several key mechanisms: dual Q-networks to reduce overestimation bias, target policy smoothing to reduce variance in policy updates, delayed policy updates to improve stability, and soft target updates with coefficient $\tau = 0.005$.

The training process begins with network initialization (lines 3-4), where both main and target networks are initialized. The main loop consists of episode-level iteration (lines 5-25) and step-level iteration within each episode (lines 6-24). During each step, the agent selects actions using the current policy with exploration noise (line 7), where $\mathcal{N}(0, \sigma_{explore})$ adds Gaussian noise for exploration.

The environment step (line 8) executes the action and returns the next state, reward, and termination signal. The transition is stored in the experience replay buffer (line 9). When sufficient samples are available (line 11), learning begins with batch sampling (line 12) and target action computation (line 13), where target policy smoothing reduces variance by adding clipped noise.

The temporal difference target (line 14) uses the minimum of two Q-values to reduce overestimation bias. Critic networks are updated by minimizing TD error (line 15), while the Actor network is updated less frequently (lines 17-18) to improve stability. The delayed policy update mechanism prevents rapid policy changes that could destabilize training.

\vspace{-0.5cm}
\section{Performance Evaluations}
\label{section:5}
\vspace{-0.2cm}
We evaluate TD3-Sched against three baseline approaches in a cloud-edge computing environment. The experimental setup consists of a 8-node  Kubernetes cluster running the Sock Shop\footnote{https://github.com/ocp-power-demos/sock-shop-demo} microservices application with Alibaba realistic traces\footnote{https://github.com/alibaba/clusterdata/tree/master/cluster-trace-microservices-v2022}, with comprehensive monitoring through Prometheus and Node Exporter. Each algorithm is tested across 50 episodes with 20 steps per episode under both normal and high-load scenarios. The baselines are as follows:
\vspace{-0.2cm}
\begin{itemize}
 \item DQN \cite{zhai2025joint}: A value-based reinforcement learning algorithm that combines Q-learning with deep neural networks. 
 The algorithm employs experience replay to break correlations between consecutive experiences and target networks to stabilize training, making it suitable for discrete action spaces in resource scheduling scenarios.
 
 \item DDPG \cite{xie2024deep}: An actor-critic algorithm designed for continuous action spaces that combines the policy gradient theorem with deep learning. 
 The algorithm uses deterministic policies to handle continuous action spaces efficiently, making it particularly suitable for resource allocation problems where precise control over CPU and memory allocation is required.
 
 \item BaseK \cite{poulton2023kubernetes}: A rule-based Kubernetes scheduler that implements static resource allocation strategies based on initial resource requests and simple load balancing heuristics without learning capabilities and relies on predefined rules. 
\end{itemize}

\textbf{Experimental Design.}
Our experimental environment consists of a Kubernetes cluster with 8 worker nodes, where 4 nodes simulate edge environments with limited resources (2 CPU cores, 4GB memory) and 4 nodes represent cloud data centers with abundant resources (8 CPU cores, 16GB memory). The SockShop microservices application includes 8 services (frontend, user, cart, catalogue, shipping, orders, payment, and queue-master) deployed across the cluster, generating realistic workload patterns through simulated user requests. 

The baseline algorithms are carefully designed to represent different approaches to resource scheduling. Each algorithm is configured with specific parameters to ensure fair comparison: DQN implements discrete action spaces with 10 discrete resource allocation levels for both CPU and memory, DDPG employs continuous action spaces with deterministic policies using the same network architecture as TD3-Sched but without the dual-critic mechanism and target policy smoothing, and BaseK operates as a rule-based scheduler with static resource allocation strategies.

To remove the randomness, each experiment has been repeated 4 times.  Each algorithm is trained for 50 episodes with 20 steps per episode, where each step represents a 30-second time window for resource allocation decisions. The training process uses the same reward function structure but with algorithm-specific modifications: DQN discretizes the continuous reward space, DDPG uses the original continuous reward, and BaseK operates without reward feedback. All algorithms are evaluated under both normal load (100 requests/second) and high-load (300 requests/second) scenarios to assess performance robustness. 

\begin{figure}[t]
\centering
\centering
\includegraphics[width=0.55\textwidth]{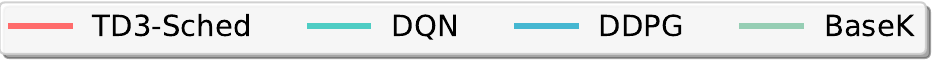}

\vspace{-0.2em} 
\begin{subfigure}[b]{0.24\textwidth}
\centering
\includegraphics[width=\textwidth]{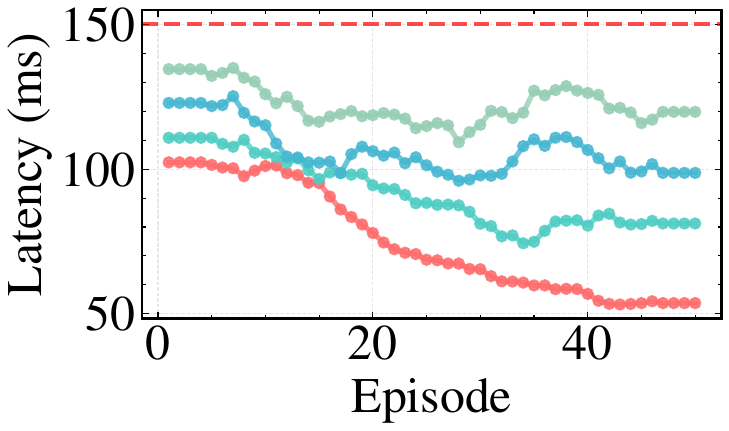}
\caption{Latency Learning in normal scenarios}
\end{subfigure}
\hfill
\begin{subfigure}[b]{0.24\textwidth}
\centering
\includegraphics[width=\textwidth]{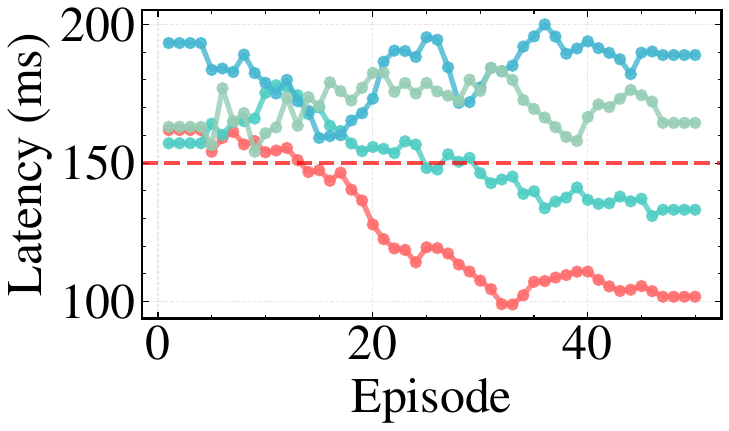}
\caption{Latency Learning in high load scenarios}
\end{subfigure}
\hfill
\begin{subfigure}[b]{0.24\textwidth}
\centering
\includegraphics[width=\textwidth]{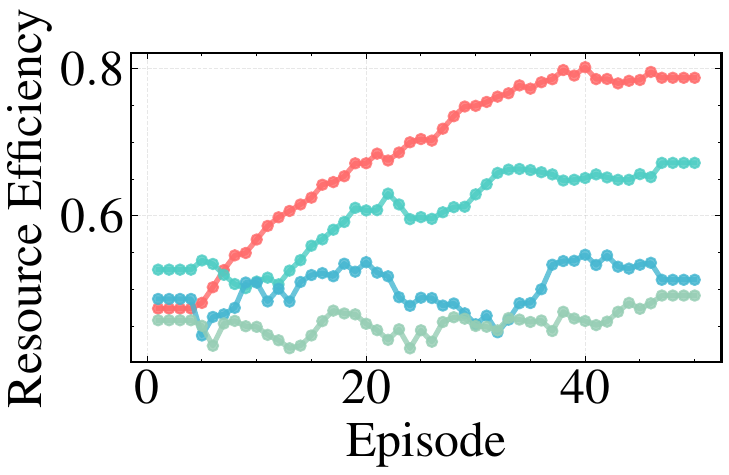}
\caption{Efficiency Learning in normal scenarios}
\end{subfigure}
\hfill
\begin{subfigure}[b]{0.24\textwidth}
\centering
\includegraphics[width=\textwidth]{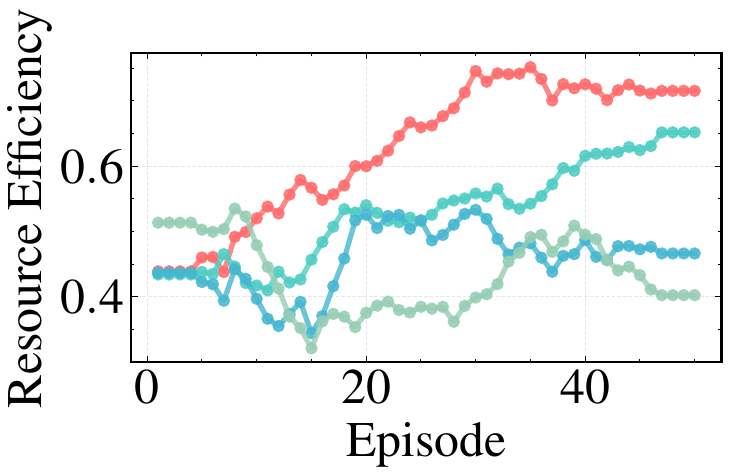}
\caption{Efficiency Learning in high load scenarios}
\end{subfigure}
\caption{Learning curves for latency and resource efficiency. TD3-Sched demonstrates faster convergence and more stable performance improvement.}
\label{fig:learning_curves}
\vspace{-0.4cm}
\end{figure}

\begin{figure}[t]
\centering
\begin{subfigure}[b]{0.24\textwidth}
\centering
\includegraphics[width=\textwidth]{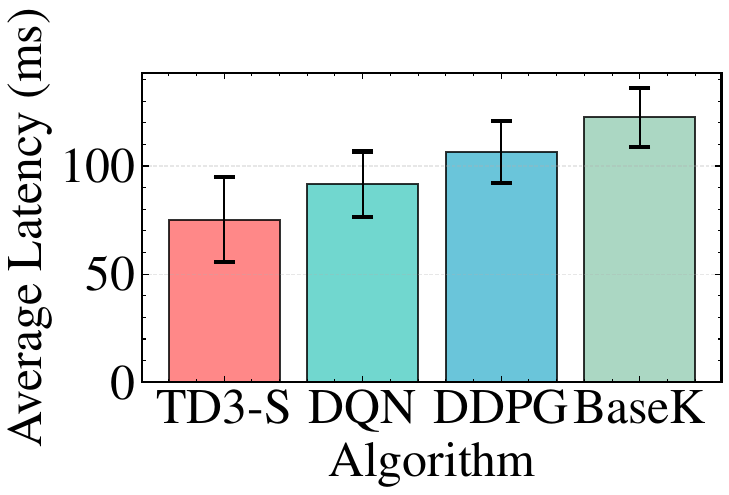}
\caption{Latency Compare in normal scenarios}
\end{subfigure}
\hfill
\begin{subfigure}[b]{0.24\textwidth}
\centering
\includegraphics[width=\textwidth]{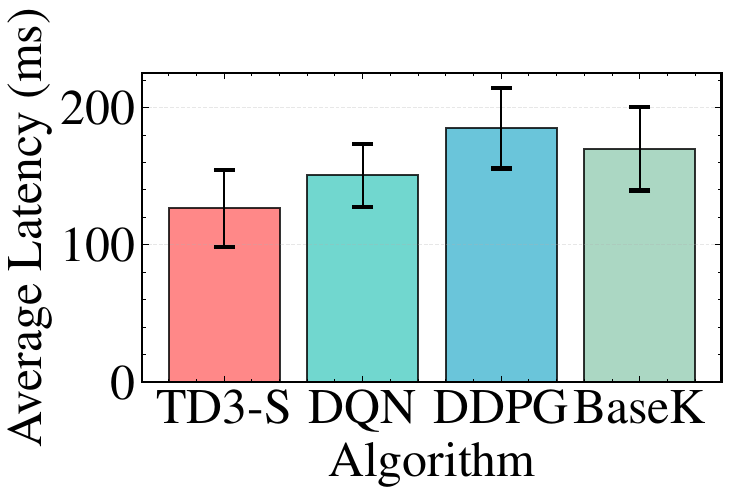}
\caption{Latency Compare in high load scenarios}
\end{subfigure}
\hfill
\begin{subfigure}[b]{0.24\textwidth}
\centering
\includegraphics[width=\textwidth]{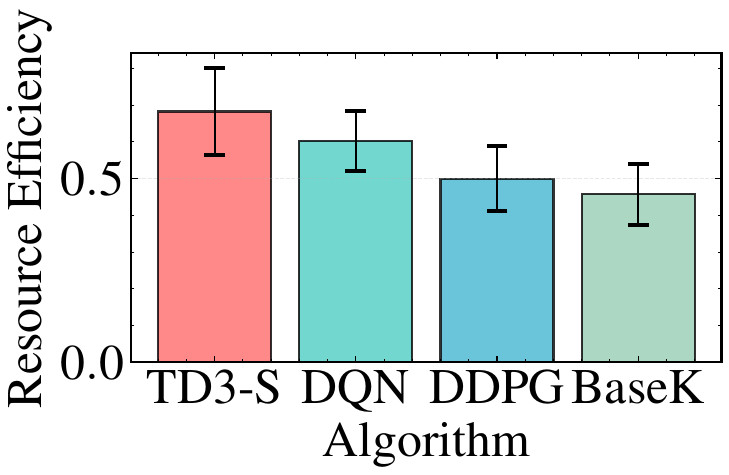}
\caption{Resource Efficiency in normal scenarios}
\end{subfigure}
\hfill
\begin{subfigure}[b]{0.24\textwidth}
\centering
\includegraphics[width=\textwidth]{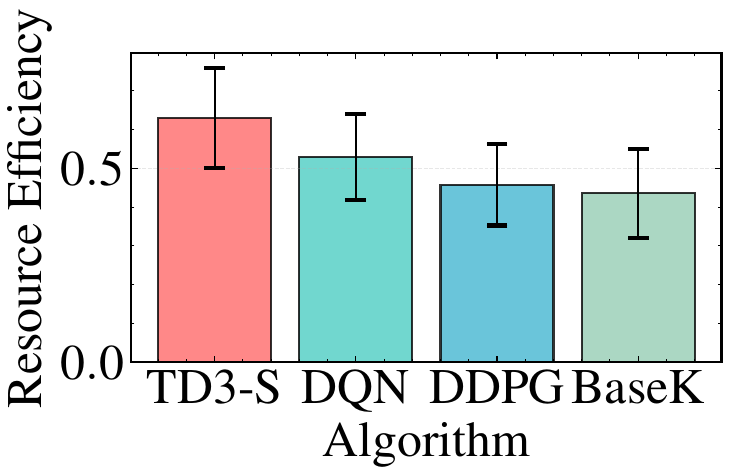}
\caption{Resource Efficiency in high load scenarios}
\end{subfigure}
\caption{Latency and resource efficiency comparison under normal and high load scenarios. TD3-Sched achieves superior performance across both metrics.}
\label{fig:latency_efficiency}
\vspace{-0.6cm}
\end{figure}
\textbf{Performance Analysis.}
Latency is measured as HTTP response time with a target of 150ms for SLO compliance. The latency calculation follows: $Latency = \frac{1}{N} \sum_{i=1}^{N} l_i^t$, where $l_i^t$ represents the response time of service $i$ at time step $t$. Under normal load, TD3-Sched achieves an average latency of 75.2ms, significantly outperforming DQN (91.6ms), DDPG (106.4ms), and BaseK (122.5ms). This improvement represents a 17.9\% reduction compared to DQN and 38.6\% reduction compared to BaseK. Under high-load scenarios, TD3-Sched maintains superior performance with an average latency of 126.6ms, while DQN, DDPG, and BaseK achieve 150.7ms, 185.0ms, and 169.9ms respectively.

Figure \ref{fig:learning_curves} presents the learning curves for latency and resource efficiency across different load scenarios. TD3-Sched demonstrates faster convergence and more stable performance improvement compared to baseline approaches. The learning curves show that TD3-Sched achieves lower latency and higher efficiency more quickly, indicating superior learning dynamics and optimization capabilities.

Figure \ref{fig:latency_efficiency} shows the comprehensive comparison of latency and resource efficiency under normal and high load scenarios. The results clearly demonstrate TD3-Sched's superior performance across both metrics, with consistent improvements over baseline approaches regardless of workload conditions.

Resource efficiency measures the balance between performance and resource utilization, calculated as: $E_{resource} = \frac{1}{N} \sum_{i=1}^{N} \frac{cpu_i^t + mem_i^t}{2}$, where $cpu_i^t$ and $mem_i^t$ represent normalized CPU and memory utilization for service $i$. TD3-Sched achieves a resource efficiency score of 0.68 under normal load and 0.63 under high load, demonstrating optimal balance between performance and resource utilization. Under normal load, TD3-Sched shows 13.3\% higher efficiency than DQN and 48.6\% higher than BaseK.

SLO violation rate is calculated as: $SLO_{violations} = \frac{1}{T} \sum_{t=1}^{T} \mathbb{I}(l_t > 150ms)$, where $\mathbb{I}(\cdot)$ is the indicator function and $l_t$ is the average latency at time step $t$. Under normal load conditions, all algorithms achieve zero SLO violations. However, under high-load scenarios, significant differences emerge in service quality control. TD3-Sched maintains excellent SLO compliance with only 0.47\% violation rate, while DQN shows 1.80\% violation rate, BaseK shows 4.42\% violation rate, and DDPG shows the highest violation rate at 5.44\%.

The reward function balances four competing objectives as defined in Section 4. TD3-Sched achieves an average reward of 62.2 under normal load and 55.9 under high load, significantly outperforming baseline approaches. Under normal load, TD3-Sched shows 28.6\% higher reward than DQN and 110.1\% higher than BaseK, demonstrating its superior learning capabilities.

\begin{figure}[t]
\centering
\begin{subfigure}[b]{0.24\textwidth}
\centering
\includegraphics[width=\textwidth]{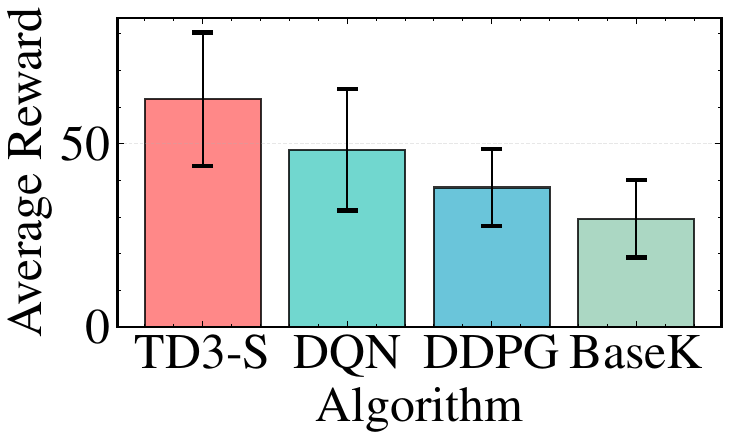}
\caption{Reward Compare in normal scenarios}
\end{subfigure}
\hfill
\begin{subfigure}[b]{0.24\textwidth}
\centering
\includegraphics[width=\textwidth]{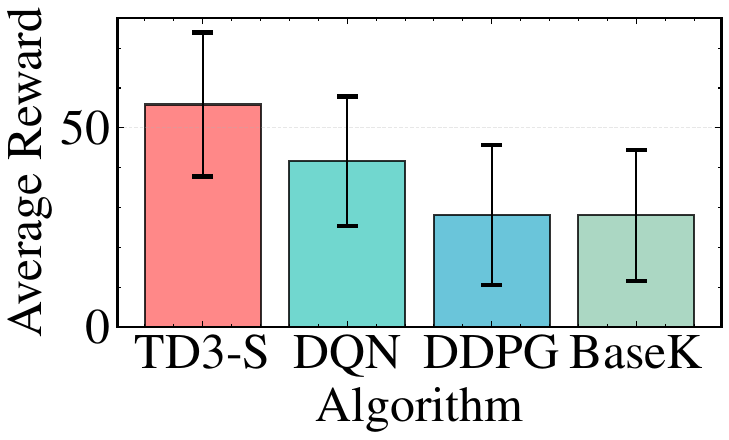}
\caption{Reward Compare in high load scenarios}
\end{subfigure}
\hfill
\begin{subfigure}[b]{0.24\textwidth}
\centering
\includegraphics[width=\textwidth]{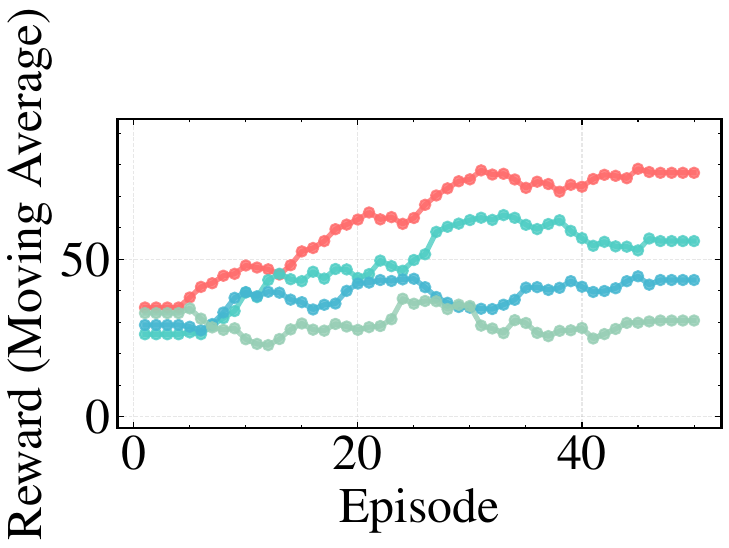}
\caption{Overall Reward Learning in normal scenarios}
\end{subfigure}
\hfill
\begin{subfigure}[b]{0.24\textwidth}
\centering
\includegraphics[width=\textwidth]{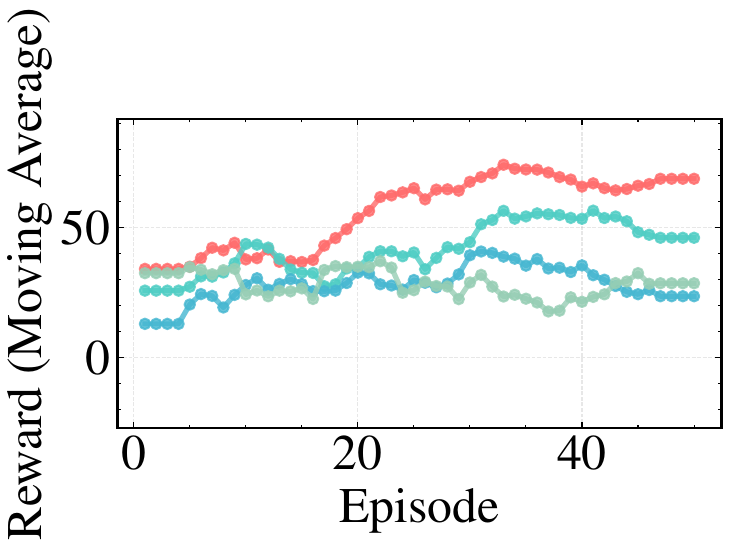}
\caption{Overall Reward Learning in high load scenarios}
\end{subfigure}
\caption{Reward comparison and overall learning curves. TD3-Sched achieves the highest average reward and fastest learning convergence.}
\label{fig:reward_learning}
\vspace{-0.7cm}
\end{figure}

Figure \ref{fig:reward_learning} illustrates the reward comparison and overall learning curves across different scenarios. The results show that TD3-Sched consistently achieves the highest average reward and demonstrates the fastest learning convergence, validating the effectiveness of the proposed approach.

The evaluation results demonstrate that TD3-Sched effectively addresses the challenges of cloud-edge resource scheduling, providing superior performance across multiple dimensions while maintaining system stability and service quality. The algorithm's robust performance under varying load conditions and its ability to maintain SLO compliance make it a compelling solution for practical cloud-edge deployments. The superior performance can be attributed to several key factors: the dual Q-network architecture effectively reduces overestimation bias that plagues single-critic approaches like DDPG, leading to more stable and accurate value estimation; the target policy smoothing mechanism prevents the policy from exploiting errors in the value function, resulting in more conservative and stable resource allocation decisions; the comprehensive multi-objective reward function enables TD3-Sched to balance competing objectives more effectively than rule-based approaches like BaseK, which lack adaptive learning capabilities; and the continuous action space allows for fine-grained resource control compared to discrete approaches like DQN, enabling more precise optimization of resource allocation in dynamic cloud-edge environments.

\vspace{-0.5cm}
\section{Conclusion and Future Work}
\label{section:7}

\vspace{-0.2cm}
This paper presents TD3-Sched, a DRL-based resource orchestration system for containers in cloud-edge computing environments. By integrating the TD3 algorithm with Kubernetes resource management, TD3-Sched achieves superior performance across multiple dimensions. Experimental evaluation demonstrates TD3-Sched's effectiveness, achieving 75.2 ms average latency under normal load (17.9\% and 38.6\% reductions compared to DQN and BaseK respectively) and maintaining excellent SLO compliance with only 0.47\% violation rate under high-load conditions. The results validate that reinforcement learning can effectively address the complex challenges of cloud-edge resource scheduling.
While TD3-Sched demonstrates significant improvements over existing approaches, several promising directions for future research emerge. 

Future work could explore adaptive weight adjustment mechanisms for the reward function, multi-tenant support with different service requirements, extension to additional resource types such as GPU resources and network bandwidth, online learning capabilities for continuous adaptation, and large-scale real-world deployments. The success of TD3-Sched opens new possibilities for applying advanced reinforcement learning techniques to cloud-edge computing challenges, providing a solid foundation for future research in intelligent, adaptive resource management systems that are essential for the future of cloud-edge computing. 
\vspace{-0.2cm}

\begin{credits}
\subsubsection{\ackname}This work is supported by National Natural Science Foundation of China under Grant 62572462, National Key Laboratory of Electromagnetic Energy (No.  614221723040201), Guangdong Basic and Applied Basic Research Foundation (No. 2024A1515010251, 2023B1515130002), Key Research and Development and Technology Transfer Program of Inner Mongolia Autonomous Region (2025YFHH0110) and Shenzhen Science and Technology Program under Grant JCYJ20240813155810014.
\end{credits}
\color{black}
\newpage

\printbibliography
\end{document}